\documentclass[journal]{IEEEtran}
\usepackage{acronym}
\usepackage{amsfonts}
\usepackage{amsmath}
\usepackage{amssymb}
\usepackage{amsthm}
\usepackage{bm}
\usepackage{cite}
\usepackage{enumerate}
\usepackage{dsfont}
\usepackage{graphicx}
\usepackage{mathtools}
\usepackage{nicefrac}
\usepackage[caption=false,font=footnotesize]{subfig}

\usepackage{xspace}
\usepackage{xcolor}
\acrodef{ML}[ML]{Machine Learning}
\acrodef{IDS}[IDS]{Intrusion Detection System}
\acrodef{IoT}[IoT]{Internet of Things}
\acrodef{QBC}[QBC]{Query-by-Committee}
\acrodef{SVM}[SVM]{Support Vector Machine}
\acrodef{NN}[NN]{Neural Network}
\acrodef{KNN}[KNN]{k-Nearest Neighbor}
\acrodef{EGL}[EGL]{Expected Gradient Length}

\graphicspath{ {./} }

\title{Active Learning for Wireless IoT Intrusion Detection}

\author{Kai~Yang,~Jie Ren,~Yanqiao~Zhu, and Weiyi Zhang%

\IEEEcompsocitemizethanks{ 
\IEEEcompsocthanksitem Accepted by IEEE Wireless Communications.

\IEEEcompsocthanksitem Kai Yang, Jie Ren, and Yanqiao Zhu are with the Department of Computer Science, Tongji University \protect\\
E-mail: \ {kaiyang@tongji.edu.cn} %

\IEEEcompsocthanksitem Weiyi Zhang is with AT\&T Labs Research\protect\\
E-mail: \ {maxzhang@research.att.com}

\IEEEcompsocthanksitem \copyright 2018 IEEE. Personal use of this material is permitted. Permission from IEEE must be obtained for all other uses, in any current or future media, including reprinting/republishing this material for advertising or promotional purposes, creating new collective works, for resale or redistribution to servers or lists, or reuse of any copyrighted component of this work in other works.
} %
\showboxdepth=\maxdimen
}
\begin{document}
\maketitle
\begin{abstract}
\ac{IoT} is becoming truly ubiquitous in our everyday life,  but it also faces unique security challenges. Intrusion detection is critical for the security and safety of a wireless IoT network. This paper discusses the human-in-the-loop active learning approach for wireless intrusion detection. We first present the fundamental challenges against the design of a successful \ac{IDS} for wireless IoT network. We then briefly review the rudimentary concepts of active learning and propose its employment in the diverse applications of wireless intrusion detection. Experimental example is also presented to show the significant performance improvement of the active learning method over traditional supervised learning approach. While machine learning techniques have been widely employed for intrusion detection, the application of human-in-the-loop machine learning that leverages both machine and human intelligence to intrusion detection of IoT is still in its infancy. We hope this article can assist the readers in understanding the key concepts of active learning and spur further research in this area.
\end{abstract}
\begin{IEEEkeywords}
	Internet of things, intrusion detection, active learning, human-in-the-loop machine learning
\end{IEEEkeywords}

\section{Introduction}
\label{sec:introduction}
\ac{IoT} refers to a network of connected physical devices that are embedded with software, sensors, electronics, and communication modules. The IoT allows the physical objects to be sensed and controlled remotely over network infrastructure and consequently enable direct integration of the physical world with computing devices. The IoT can help to improve the efficiency, reliability, and accuracy of existing systems and has received significant research interest recently.

Wireless communications techniques such as NB-IoT, WiFi, Bluetooth Low Energy have become the de facto standard for connecting IoT devices due to its flexibility, low cost, and simple installation and maintenance processes. These wireless devices can be simply installed in offices or factories to provide seamless connectivity. In addition, the commercial 5G network is anticipated to be launched by the end of 2020, which together with LTE, LTE-A, and WiFi, could provide high-rate coverage as well as harmonious quality of service for every user.

Several papers (e.g., \cite{Cheng2016,Hei2013,Wu2016}) have studied related  \ac{IoT} security issues.
In spite of the low cost and flexibility, the \ac{IoT} come with unique vulnerabilities than traditional networks and consequently face a variety of security challenges, such as the DoS attacks.
Intrusion Detection System (IDS) plays a key role in detecting network attacks. However, the \ac{IoT} have some major characteristics such as the sensor nodes are usually power-limited, have limited memory space, and the capacity of wireless channels is very limited. In addition, each node in IoT has an IP address so that any user can interact with the IoT node from anywhere in the world, which makes it particularly vulnerable to cyber security attacks. Such challenges make the design of the \ac{IoT} \ac{IDS} different from the traditional network \ac{IDS}.

Intrusion detection technologies can be broadly grouped into three categories, namely the misuse-based methods, the anomaly-based methods, and the hybrid methods \cite{Kai2017,Zarpel2017}. The misuse-based method first constructs a collection of signatures based on information such as domain knowledge and expert experience. It then tries to look for a particular pattern in the incoming network data that closely matches one or multiple signatures in the database. The misuse-based method can effectively detect intrusions matching at least one of the signatures in the database and thus has low false-positive rate. However, it cannot detect unknown intrusions that do not match any pattern in the database, and consequently may give rise to high false-negative rate, especially when the attacker is aware of signatures in the database. As a remedy, the misuse-based IDS often needs to frequently update the signatures and rules in the database.

The anomaly-based method first learns the normal network behaviors and then identifies anomalies that do not comply with the normal network conditions. The anomaly-based method can effectively detect unknown attacks that do not happen before. In addition, since the normal network behaviors are learned by the machine and in most cases no explicit rules for intrusion detection are provided, it is less likely for the attackers to learn the rules and make their attacking strategies undetectable. The main disadvantage of the anomaly-based method is that it may generate a huge volume of false alarms because any previously unseen behaviors can be treated as anomalies.

The hybrid approach aims to combine the misuse-based and anomaly-based methods and thereby achieve the advantages of both. In reality, most IDS systems employ a hybrid approach to strike a balance between the false-positive rate and false-negative rate.

One goal of machine learning is to let computers learn from and make predictions based on observed data without being explicitly programmed. While machine learning techniques have been extensively studied for anomaly detection, their applications to intrusion detection in practice are relatively limited. This is partly because the design of intrusion detection system in practice often lacks sufficient training data and also requires domain knowledge of the particular system under investigation.

Active learning is a subfield of machine learning that emphasizes on learning from limited amount of training samples \cite{Settles09}. It is naturally suited for the design of \ac{IDS} since providing labels for intrusion detection is usually time-consuming. In some cases, labeling is impossible for intrusion that never happens before. In short, active learning can harness the power of machine learning together with the experience from domain expert. Consequently, it can significantly decrease the labeling efforts and quickly build a machine learning model for intrusion detection. In addition, the active learning framework allows quick update of the machine learning model and therefore can be updated in a short period of time for new network attacks.

The remainder of this paper is organized as follows. Section II is devoted to the intrusion detection of wireless IoT network. In Section III, we overview the rudimentary concepts of active learning and the query strategies. An example of applying active learning method to the intrusion detection problem is also provided. We conclude our discussion in Section IV.

\section{Intrusion Detection for Wireless Internet of Things}
\label{sec:IDS-for-IoT}

Intrusion detection aims at detecting harmful activities made by internal or external intruders against the system. Typical intrusions include information gathering, eavesdropping, harmful packet forwarding, packet dropping, hole attacks etc.
To detect these intrusions, an \ac{IDS} is often designed with a powerful detection engine, a reporting module along with a group of sensors, where the sensors are deployed to gather data and monitor the system, the engine analyzes the gathered data to detect suspicious activities, and once an intrusion is detected, the reporting module will generate an alert for further neutralizing the attack \cite{Zarpel2017}.

The \ac{IoT} has some major characteristics. Firstly, unlike the wireless sensor networks, the IoT has a novel architecture that there always exists an edge node, i.e. border router, that connects the IoT network with the Internet. This architecture can be utilized in building centralized or hybrid \ac{IDS}s. Secondly, the sensor nodes in \ac{IoT} are directly connected to the untrusted Internet and globally identified by their IP addresses which make the IoT more vulnerable to intrusions from the Internet. Lastly, the \ac{IoT} nodes are resource-constrained and connected through lossy links.
How to effectively exploit these opportunities and threats makes the building of \ac{IDS} for the \ac{IoT} a very challenging work.

\begin{figure*}
		\centering
		\includegraphics[width=0.8\textwidth]{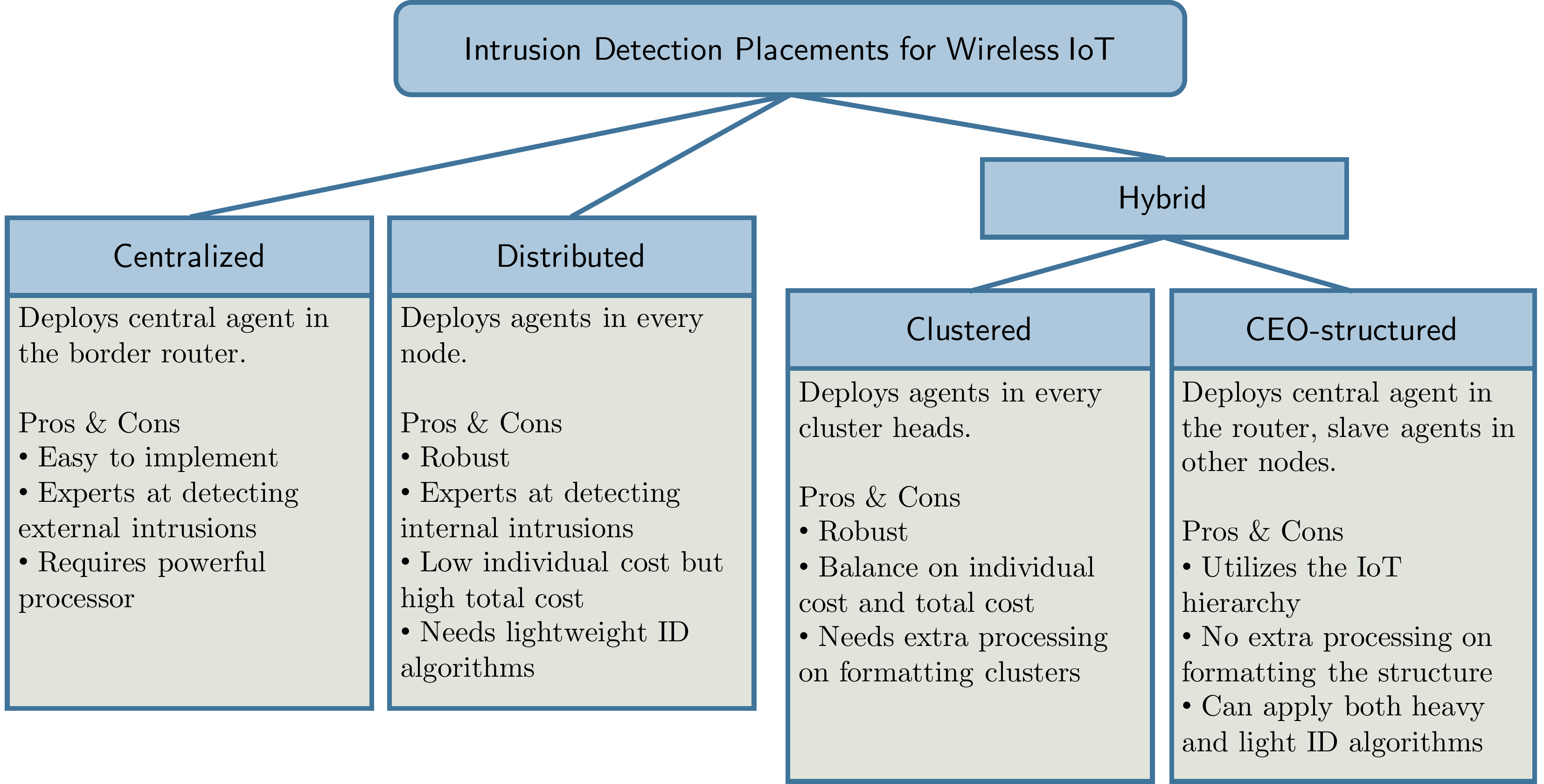}
		\caption{Placement strategies of intrusion detection.}
		\label{fig:intrusion-detection}
\end{figure*}

As shown in Figure \ref{fig:intrusion-detection}, depending on the placement strategies, the \ac{IDS}s can be classified into centralized, distributed, and hybrid, where the hybrid systems can be furthered categorized into clustered and CEO-structured.

\subsubsection{Centralized IDS for IoT}
A centralized \ac{IDS} usually deploys its agent in a master node with powerful computation resource and large storage/memory space. This master node must also have the ability to monitor all the activities in the network, so that the deployed \ac{IDS} agent can access the network activity data in real time and detect intrusions by analyzing the network activities instantaneously.

The centralized \ac{IDS}s are easy to implement in the \ac{IoT} partially because the \ac{IoT} system naturally has an edge node (i.e. border router) that connects the \ac{IoT} network with the Internet. Network intrusions from external intruders can be better detected by the centralized \ac{IDS} agent since all outside packets have to be sent to the edge node, and the attack can be mitigated by simply dropping the harmful packets sending from the malicious intruders at the edge node. On the other hand, detecting intrusions from internal intruders might be hard for centralized \ac{IDS}s since it requires the \ac{IDS} agent to deeply monitor actions of resource limited sensor nodes through lossy networks. Meanwhile, although border routers usually have a more powerful processor than the sensor nodes connected to it, computation cost, energy consumption, and  memory usage are still the top considerations when the \ac{IDS}s are installed in the \ac{IoT}. Besides the strength in detecting external intrusions, the centralized \ac{IDS}s are also capable of detecting some internal attacks such as the selective forwarding attack, in which malicious nodes selectively forward packets to disrupt routing paths.

\subsubsection{Distributed \ac{IDS} for \ac{IoT}}

Unlike the centralized scheme, a distributed \ac{IDS} deploys detection agents in every sensor node. Each agent monitors and analyzes the behaviors of its neighboring nodes within its radio range and generates alerts if it observes abnormal activities. The decision of whether or not a node is compromised can be made either by the local agent's own analysis, which is called individualized decision-making, or by the majority votes of all neighboring agents, which is called cooperative decision-making.

The distributed \ac{IDS}s have a strength in detecting internal intrusions. Besides, the deployment of the distributed \ac{IDS} does not rely on a super powerful central node, all sensor nodes with detection agents in the \ac{IoT} can work collaboratively to detect potential attacks, hence the system is more robust in the sense that accidents causing one node to malfunction will not be detrimental to the wireless IoT network as a whole. However, the distributed \ac{IDS} techniques are usually not energy-efficient because the \ac{IDS} agents are installed in every node, which increases the total computation cost and incurs extra communication cost for cooperation. In building distributed \ac{IDS}s for \ac{IoT}, the designer must pay special attention to the energy consumption on each individual agent since the detection agents will be installed on sensor nodes with limited power, memory space, and computing capability. The economic cost should also be taken into consideration because deploying detection agents will affect the battery life of the sensor nodes in the \ac{IoT}.


\subsubsection{Hybrid IDS for IoT}
By synthesizing the centralized strategy and distributed strategy, the hybrid placement strategy seeks for a tradeoff among detection accuracy, algorithm complexity, energy consumption, memory and computing power usage, communication cost, and other design requirements. The nodes being selected to host \ac{IDS} agents are often more robust and powerful. The hybrid placement strategy can be further categorized into clustered and CEO-structured.

In the clustered placement, the network is organized into clusters, where a cluster head is selected from each cluster, and deployed with the \ac{IDS} agent. For example, the cluster heads can overhear and monitor their neighbor nodes' packet transmission \cite{Amaral2014}. A specification-based detection algorithm is then applied, where the overheard packets are compared to the abnormal behaviors (set of rules) which are manually defined.

In clustered \ac{IDS}s, clustering algorithms may consume considerable amount energy through the formation of the clusters. On the contrast, the CEO-structured \ac{IDS}s, without requiring any preprocessing to classify the nodes, utilize the natural hierarchy of the \ac{IoT} to place a central processing agent often in the border router node, and all other agents in those resource-constrained sensor nodes.
The central processing agent in the border router is in charge of tasks
that demand high computation capacity and memory usage, while the agents deployed in constrained nodes are lightweight.
			
\section{Active Learning for Anomaly Detection}
\label{sec:active-learning}

\ac{ML} deals with the problem of extracting features from data to solve predictive tasks, such as classification, regression, clustering, and decision making. Classic \ac{ML} algorithms usually involve no human effort other than some data preprocessing and feature selecting work. However, the performance of applying automated \ac{ML} can be severely constrained due to the lack of clean data, meanwhile, in some learning tasks where data is limited, human experts can show very competitive results.
To overcome the difficulty of data insufficiency and reduce the long processing time in \ac{ML}, a human-in-the-loop scheme is introduced, which typically utilizes guidance of domain experts to adjust and optimize the learning behaviors in the training phase.
It is believed that bringing human experience in the \ac{ML} loop would greatly enhance the knowledge discovery process, therefore achieving better results than classic \ac{ML} algorithms.

\subsection{Active Learning}
\begin{figure}
	\centering
	\includegraphics[width=\columnwidth]{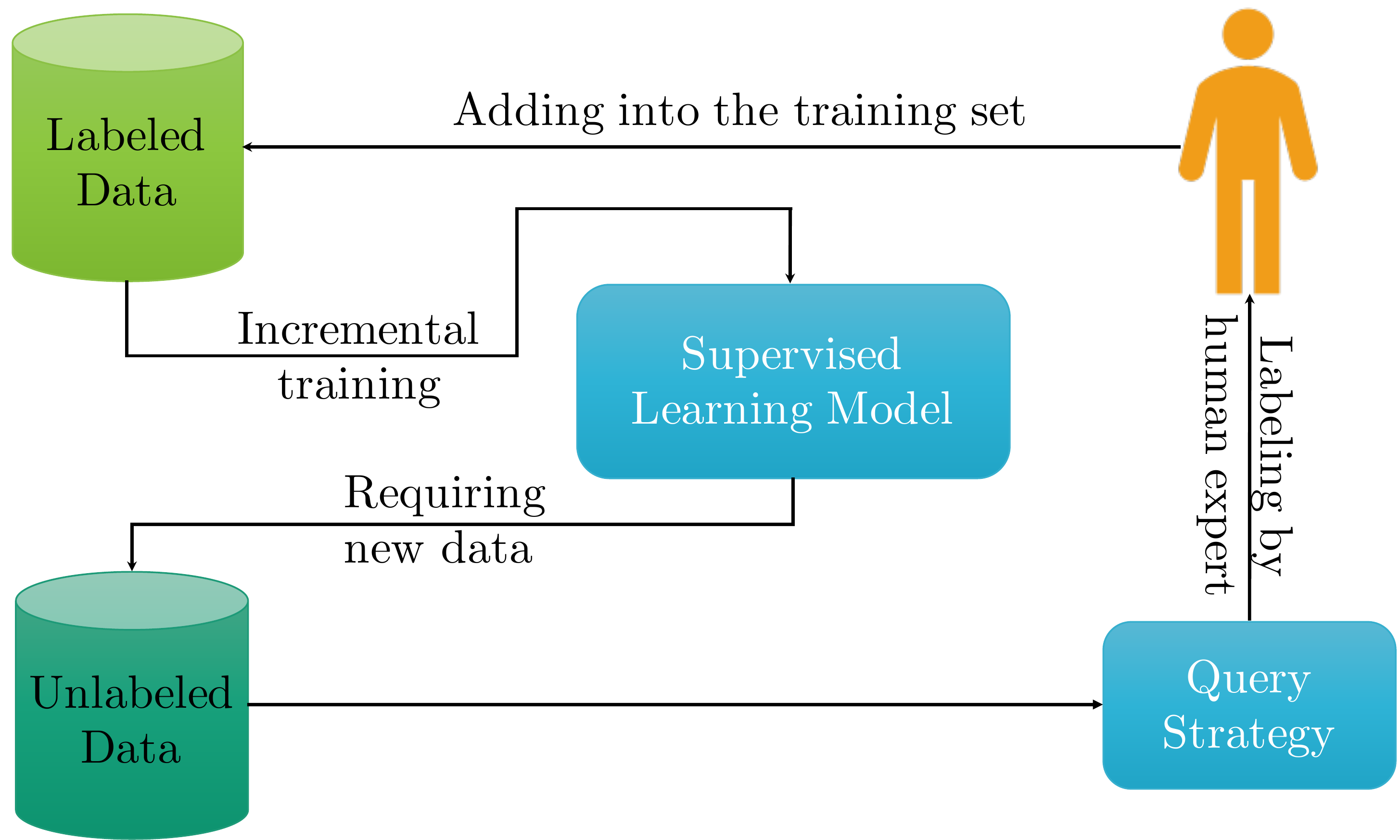}
	\caption{An illustration of active learning.}
	\label{fig:active-learning}
\end{figure}

Active learning is a subfield of human-in-the-loop machine learning where humans play the role of ``omniscient" to label the selected data. In many real learning problems, obtaining unlabeled data is much less expensive than labeled data, meanwhile, data are not evenly useful for a learning method in the sense that some of them are dirty, redundant or trivial. To address these issues, the active learning technique is proposed. It is designed to achieve high accuracy using as few labeled instances as possible, thereby minimizing the cost of obtaining labeled data\cite{Settles09}. As shown in Figure \ref{fig:active-learning}, a classical active learning system contains a select query, a human annotator and a machine learning model. The select query selects unlabeled data based on certain strategies. A human annotator, then labels the selected data and adds it into the training set. The learning model adjusts its parameters every time it receives new labeled data.
The whole learning process stops when the system achieves desired prediction accuracy.
It is showed by both empirical study and theoretical analysis that, by carefully selecting data to label, active learning can achieve same accuracy with much fewer labeled data than classic \ac{ML} algorithms.

Active learning can be classified into stream-based selective sampling scheme and pool-based sampling scheme. In stream-based selective sampling, the data will first be sampled one at a time from the actual distribution, then the learner can decide for each sampled data, whether to request its label or discard it. In pool-based sampling, all unlabeled data are gathered in a data pool, then the query engine will select a data instance from the pool and send it to the human annotator. The decision of whether or not to discard a data in the stream-based method, as well as the selection of instance in the pool-based method, usually follows certain query strategies, which all aim at accelerating the learning process.
In the remaining of this section, we will first go over some general frameworks of query strategies that follow the categorizing method of \cite{Settles09}, we will then review some works of applying active learning in building \ac{IDS}s, finally, we will give an example of applying active learning in building \ac{IoT} \ac{IDS} specifically.

\subsection{Query Strategies of Active Learning}
Typical query strategies include uncertainty sampling, query-by-committee, expected model change etc. We first give a brief review in the sequel.

\subsubsection{Uncertainty Sampling}
Motivated by results in computational learning theory, the uncertainty sampling strategy queries the instance which is least certain to label under the current model. The most uncertain samples are expected to be the most informative and are capable of improving the model most \cite{Settles09}.

When probabilistic learning models are in use, the meaning of uncertainty is very straightforward: the posterior probability of an instance belonging to a certain class represents the confidence of classifying the instance to that class, hence uncertainty sampling strategy tends to select a instance of which the learning model is unconfident to classify to any class. For binary classification problems, it is the instance with posterior probability of being positive closest to $0.5$ \cite{Lewis1994}.

More generally, the uncertainty can be evaluated under different criteria, e.g., entropy, least confident of prediction, and least margin. The entropy criterion selects the instance with the highest Shannon information entropy. The least confident of prediction selects the instance achieving the min-max posterior probability. The least margin criterion selects the one with minimum difference of posterior probabilities between the first and second most probable class. It is also possible to apply uncertainty sampling in non-probabilistic classifiers.

\subsubsection{Query-by-Committee}
The \ac{QBC} strategy, considers learning by committee setup, where each member in the committee represents a learning model or a hypothesis. The hypotheses that fit the labeled training data formed the version space. By viewing the learning problem as a search for the best hypothesis that describes the data in the version space, the active learning problem then becomes to shrink the version space with the least amount of new labeled data. For each selected data instance, the hypotheses committee will vote on the class it belongs to. The selected data instance is chosen according to the principle of maximal disagreement among the committee members \cite{Seung1992}. The committee of models will be trained on the new labeled data instances.

To measure the disagreement, the vote entropy metric and average Kullback-Leibler divergence metric are often used. The vote entropy views the voted class as a random variable with a probability of being class $c$ equals to the portion of committee members that labeling the instance to $c$ under the current model, and then uses Shannon information entropy to measure the uncertainty of labeling this instance. The averaged KL divergence measured an average distance of each committee member's labeling to the whole committee's labeling. Instead of picking one class per committee member deterministically, a soft vote may also be applied in the QBC strategy where each member outputs its posterior label probability/confidence vectors.

\subsubsection{Expected Model Change}
The expected model change strategy uses a decision-theoretic approach. Under this query framework, the data instance that changes the current model parameters (i.e. gradient) most will be selected.
The model's change is most commonly measured by \ac{EGL}, which can be broadly used for any gradient descent learning methods \cite{Settles2008}.

For each instance, new gradient can be computed after the instance is labeled and added into the training set. However, since the true label is not aware in the query phase, the expectation of the new gradient over all possible labels will be computed. The data instance with the largest expected gradient will be selected. The expected model change strategy prefers instance that has the largest influence on the model, hence could avoid labeling redundant data. However, the major drawback is that, the feature space in real problems may not be isotropic, hence an ill-conditioned feature component may severely decrease the system performance.

\subsubsection{Expected Error Reduction}
The expected error reduction strategy also uses a decision-theoretic approach, measures how much its generalization error will be reduced. Under this query strategy, a model will first be trained for each data instance along with the existed labeled training set, then the expected error of the remained unlabeled instances tested in this model will be calculated. The data instance with the minimized expected loss will be selected. The $0/1$-loss metric and the log-loss metric are usually considered. The expected error reduction strategy with a log-loss metric can also be interpreted as seeking for data instance that minimizing the expected entropy over the unlabeled set or maximizing the expected information gain of the query. The major constraint of using the expected error reduction strategy is the computation cost in the process of model re-training and error estimation for each data instance.

\subsubsection{Variance Reduction}
Due to the huge computational cost of model re-training and error estimation, it is hard to minimize the expected error directly, yet it is possible to reduce the expected error indirectly by minimizing the output variance, which for some scenarios, can be formalized to a differentiable function. For these scenarios, gradient method can be used to find the best query instead of testing all unlabeled instances, which could speed up the learning process significantly.

\subsubsection{Information Density}
Outliers are very easy to be lead into the training set in the querying frameworks such as uncertainty sampling, \ac{QBC}, and \ac{EGL}. The reason is that these strategies are only interested in selecting the less certain instances, however, the selected instances may lie on the tail of the distribution hence are not representative to the other instances.

To solve this problem, the information density is introduced to modify the base querying strategy.
In the density strategy, the evaluated metric of the base querying strategy will be modified by the data instance representativeness, where the representativeness is measured by the closeness of this data instance to all other instances. By using the density strategy, the system will select those instances with a balance on informativeness and representativeness.

\subsection{Active Learning in Intrusion Detection}

In machine-learning-based intrusion detection techniques, detection models are trained from the training data set which includes both attack data and normal data. These models must be updated periodically, in order to improve the intrusion detection performance and recognize new attack types on the basis of the previous results. However, the major drawback of applying machine learning in \ac{IDS} is that it usually requires a large amount of labeled training data whereas in many of the intrusion detection scenarios, obtaining adequate attack data is always time-consuming and greatly relies on the domain experts. This is particularly challenging for designing IDS for wireless IoTs since IoT devices are of limited power, memory space, and computing capabilities. Furthermore, the limited wireless channel capacities between various wireless IoT nodes render the collection of a large amount of training data impossible. It is therefore especially difficult to collect a large amount of intrusion data from wireless IoT network.

By selecting the most ``useful" unlabeled data in the querying phase, active learning effectively integrates the power of machine learning and the experience of domain expert, greatly reduces the labeling cost, and significantly accelerates the training process, which helps the machine-learning-based \ac{IDS} better meet the design requirements such as resource limitation, detection response time, and system updating period.

\begin{figure*}
    \centering
  \subfloat[]{
       \includegraphics[width=0.45\textwidth]{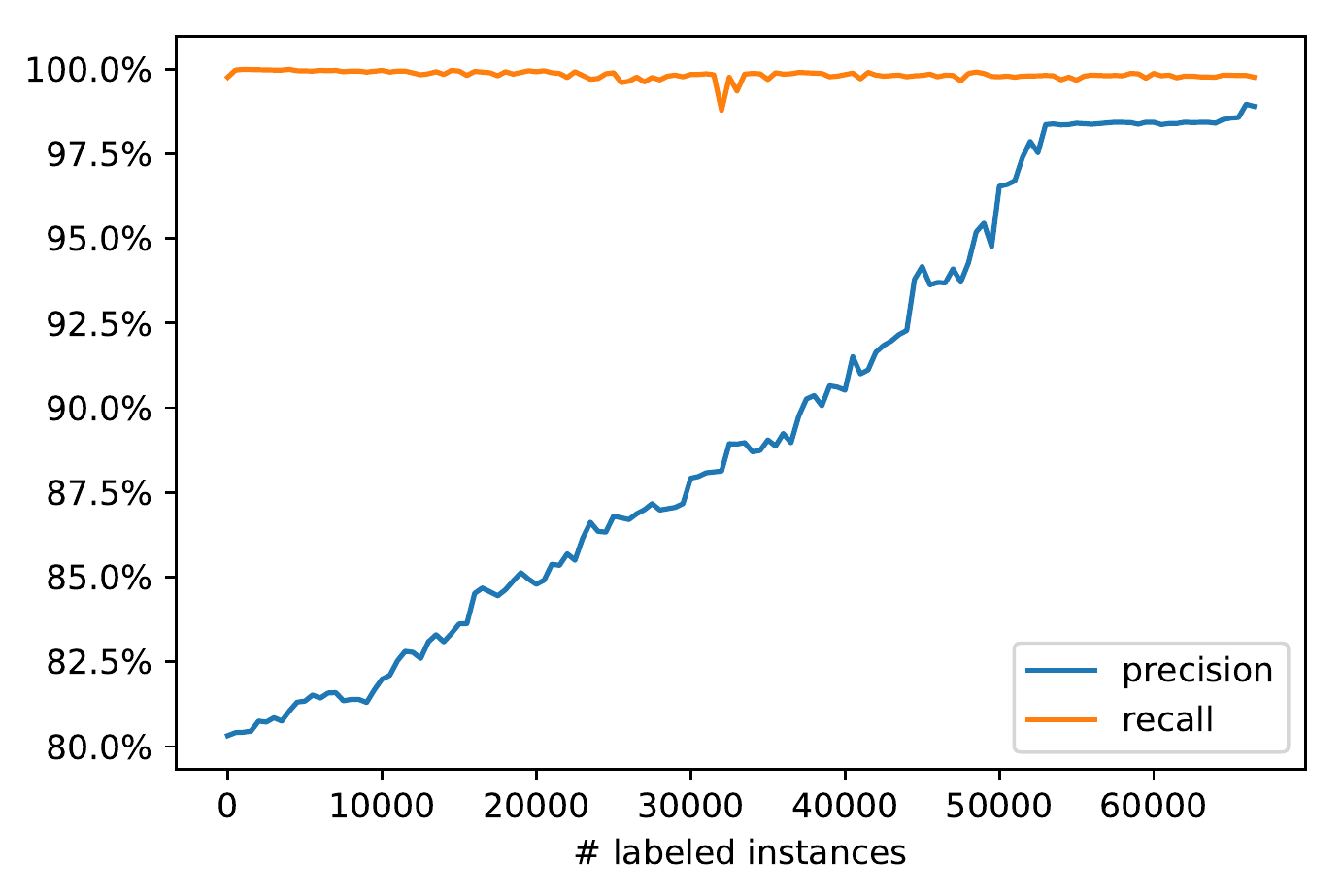}}
    \label{fig:kdd99-active-learning}\hfill
  \subfloat[]{
        \includegraphics[width=0.45\textwidth]{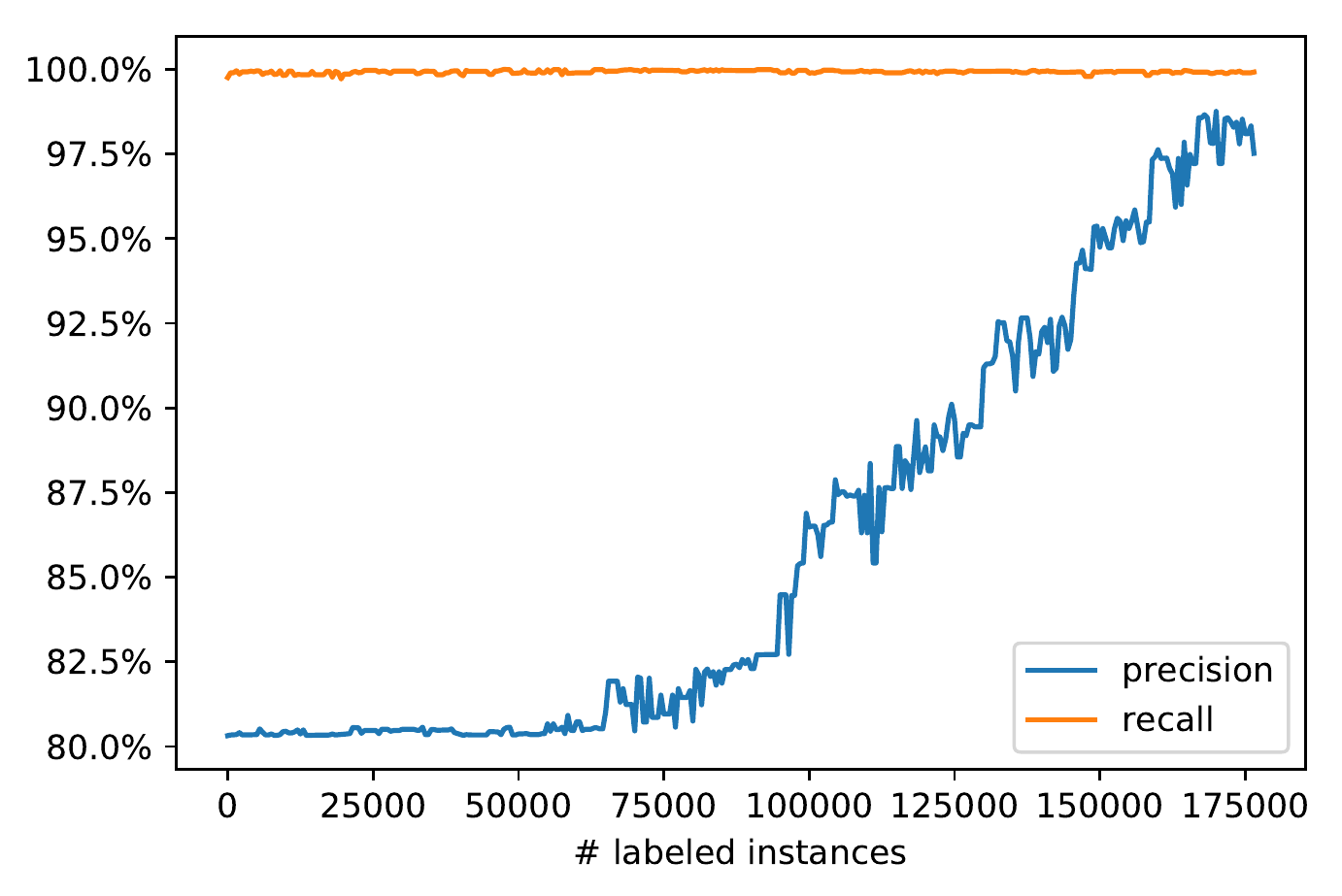}}
     \label{fig:kdd99-random} \\
  \caption{Experiment results (KDD 99 Dataset) of using (a) the active learning method, and (b) the random selection method.}
  \label{fig:kdd99-results}
\end{figure*}

\begin{figure*}
    \centering
  \subfloat[]{
       \includegraphics[width=0.45\textwidth]{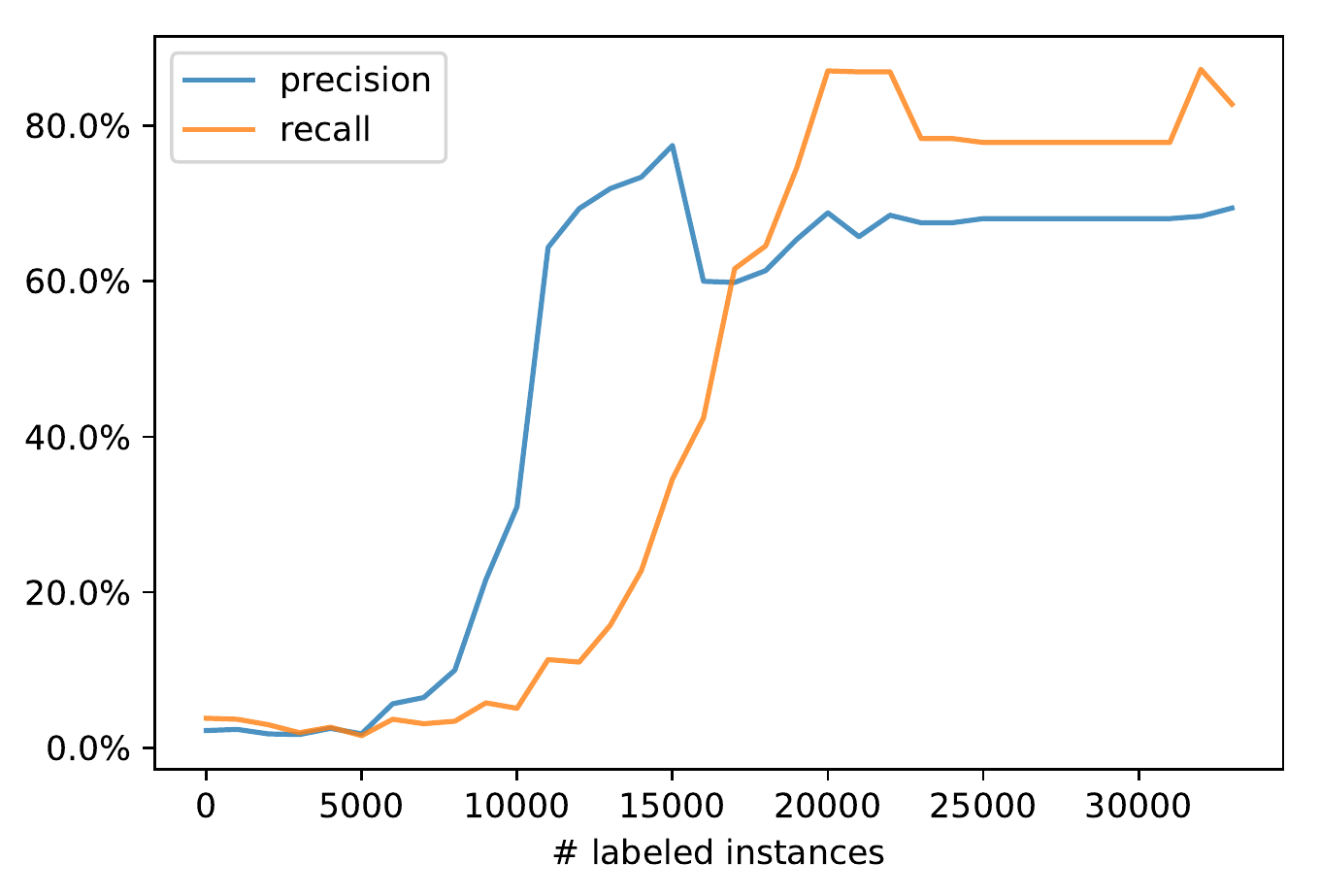}}
    \label{fig:AWID-active-learning}\hfill
  \subfloat[]{
        \includegraphics[width=0.45\textwidth]{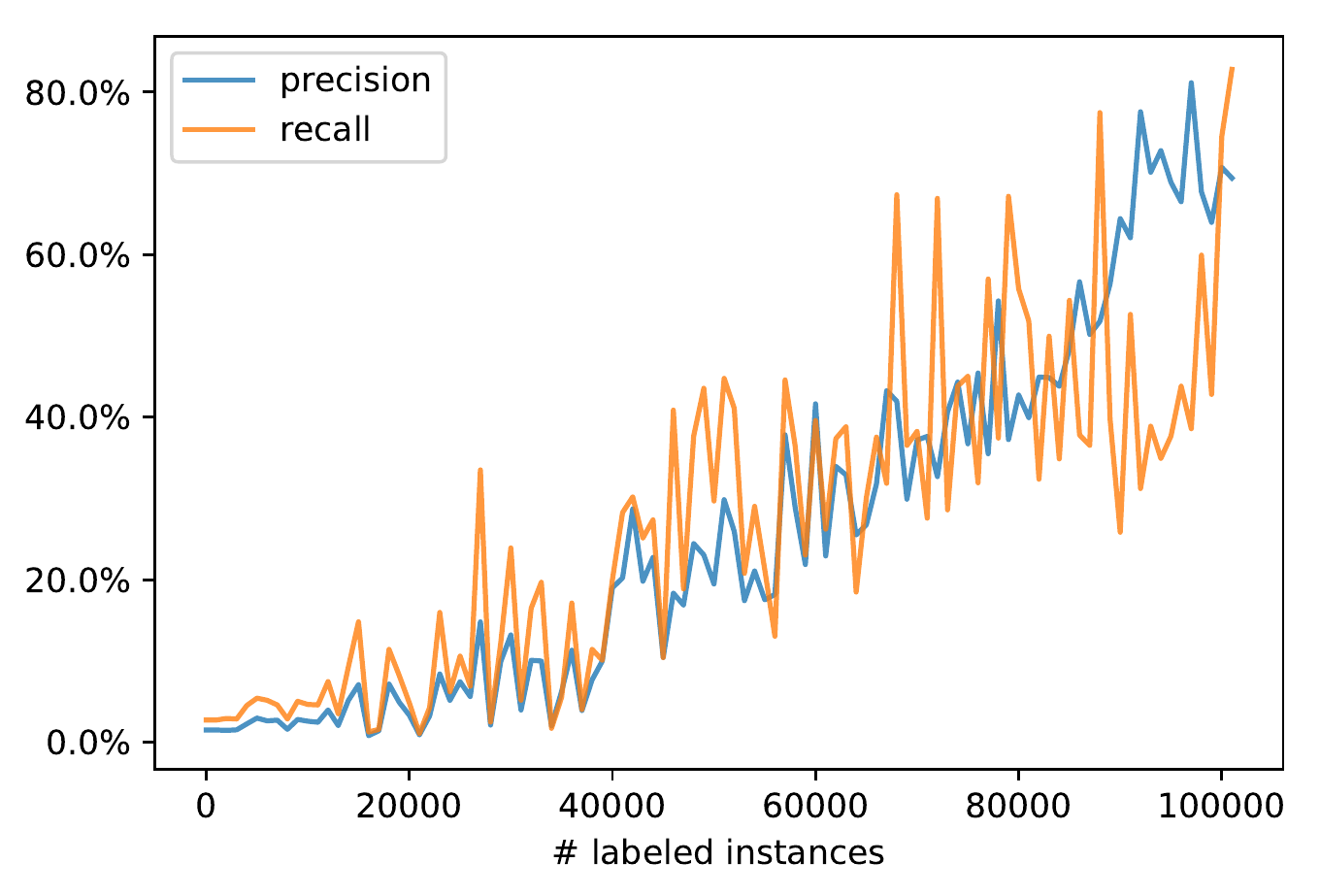}}
     \label{fig:AWID-random} \\
  \caption{Experiment results (AWID Dataset) of using (a) the active learning method, and (b) the random selection method.}
  \label{fig:AWID-results}
\end{figure*}

Consider the structural simplicity and computational efficiency, the uncertainty sampling is the most commonly used querying framework in active-learning-based \ac{IDS}s. In the uncertainty sampling strategy, the \ac{IDS} selects the data instance that the system is least confident about, and hence requires the human expert to help labeling. This data instance, after being labeled by a human expert, will be added into the training set to update the \ac{IDS}.

For probabilistic learning models, the uncertainty can be measured under many criteria, such as entropy, least confident of prediction, and least margin. For binary classification problems, the system seeks for a data instance whose posterior probability of being attacked is nearest to $0.5$. Non-probabilistic models such as \ac{SVM} and \ac{NN} can also be utilized for intrusion detection problems \cite{Almgren2004,Li2007}.

Although uncertainty sampling shows great promise in active learning based \ac{IDS} due to its simple structure and low computational cost, other query strategies such as \ac{QBC} and information density can also be considered. For example, the diversity query strategy can be used along with the uncertainty sampling in payload-based anomaly detection problems \cite{Gornitz2009}.

In the sequel, it is seen that by only selectively labeling a small portion of the entire data set, the detection rate can be significantly increased.

\subsection{Active Learning for Wireless IoT Intrusion Detection}

Although many machine learning techniques have been used in building \ac{IDS}s, there exist limited amount of literatures related to active learning for intrusion detections, among which most works employ uncertainty sampling as the selected criterion. To the best of our knowledge, no prior work discusses the active learning in building IoT \ac{IDS}s.

We next present an example of employing active learning approach for intrusion detection. In this active-learning-based method, we first employ an unsupervised local outlier factor method to detect anomalies in the data set. Then the active learning algorithm is applied. The active learning algorithm iteratively runs three steps, i.e., supervised learning, label selection, labeling by the expert labeler, until it reaches a threshold of performance in terms of precision and recall. In this case, we set precision and recall both greater than $99\%$ as the exiting condition, and employ XGBoost \cite{Chen:2016:XST:2939672.2939785} and the uncertainly sampling as the supervised learning algorithm and the label selection criterion respectively. XGBoost is an efficient extreme gradient boosting algorithm that can be used for classification. Compared to other machine learning models such as neural networks, XGBoost has fewer parameters, leading to a simpler structure. Also, current open source XGBoost implementation is scalable, portable, and distributed. Therefore, it is naturally suited for IoT platforms where computation and memory resources are strictly constrained.

As shown in Figure \ref{fig:kdd99-results}, we compare the performance of the active learning for intrusion detection with the random-select method in the experiment. $80\%$ and $20\%$ of data is chosen to be the training set and the testing set, respectively. KDD 1999 dataset is used in the study. It is seen that compared with the random-select method, the proposed active-learning-based method reaches the required performance much quicker than the random one. Also, the total number of labeled instances is reduced to almost one third to achieve the same performance, which significantly reduces the labeling time and efforts.

As another example, we have studied the performance of the proposed active learning method for the AWID \cite{7041170} dataset, which is obtained in a real WiFi environment. It is seen from Figure \ref{fig:AWID-results} that the proposed method outperforms the random-select method significantly. It can quickly build a machine learning model for intrusion detection with a small portion of the labeled data.

While active learning is particularly suitable for IDSs of wireless IoTs, there exist a few challenges that remain to be addressed. Firstly, the wireless IoT devices are of limited power, memory, and computing resources. It remains unknown how to efficiently collect the training data from such a resource-constrained distributed system. Secondly, there are a number of query strategies for the active learning, but which query strategy is best suited for the design of IDS for wireless IoT network? Finally, the obtained machine learning model for intrusion detection should be updated periodically when the network behaviors evolve according to the changing environments. How can the active learning approach be adaptively incorporated in these updates?

\section{Conclusion}
\label{sec:conclusion}
This paper studies the active learning approach for intrusion detection of wireless IoT networks. Background of network intrusion detection has been provided with emphasis on the unique security challenges faced by wireless IoT networks. We have further discussed how to apply the active learning framework to intrusion detection of a wireless \ac{IoT} network. Experimental results have been presented and discussed. It is seen that the active learning method can effectively improve the performance over the traditional supervised learning techniques for intrusion detection. While active machine learning techniques have found a variety of applications in diverse areas, there exists limited amount of work on active learning for intrusion detection of wireless \ac{IoT} networks. On the other hand, the increasing amount and the complexity of network data make the labeling process tedious and time-consuming. It is therefore crucial to reduce the labeling efforts and speed up the process of training and updating the machine learning model. We hope this work on active learning for wireless intrusion detection can draw attentions from researchers in machine learning as well as wireless security and spur further research efforts in this area.

\bibliographystyle{IEEEtran}
\bibliography{sources}
\end{document}